\documentstyle[aps,prc,twocolumn,psfig]{revtex}
\newcommand{\doe}
{One of us (NKG) was supported by the
Director, Office of Energy Research,
Office of High Energy
and Nuclear Physics,
Division of Nuclear Physics,
of the U.S. Department of Energy under Contract
DE-AC03-76SF00098.}

\begin{document}
\draft
\title{Surface tension between kaon condensate and normal
nuclear matter phase}
\author{Michael B. Christiansen}
\address{Institute of Physics and Astronomy,
University of Aarhus,
DK-8000 \AA rhus C, Denmark}
\author{Norman K. Glendenning}
\address{Nuclear Science Division and
Institute for Nuclear \& Particle Astrophysics,
 Lawrence Berkeley National Laboratory, MS: 70A-319, Berkeley,
California 94720}
\author{J\"{u}rgen Schaffner-Bielich}
\address{RIKEN BNL Research Center, Brookhaven National
Laboratory, Upton,
New York 11973}
\date{\today}
\maketitle

\begin{abstract}
We calculate for the first time the surface tension and
curvature coefficient 
of a first order phase transition 
between two possible phases of cold nuclear matter, a
normal nuclear matter 
phase in equilibrium with a kaon condensed phase, at
densities a few times the 
saturation density. 
We find the surface tension is proportional
to the difference in energy density between the two phases
squared. 
Furthermore, we show the consequences for the geometrical
structures of the 
mixed phase region in a neutron star.
\end{abstract}


\section{INTRODUCTION}

The possibility of different phase transitions taking
place in the superdense 
interior of neutron stars has been the target of
considerable interest during 
the last few decades, where pion and kaon condensation as
well as quark 
deconfinement have been investigated.
But only less than a decade ago was it realized that if
the phase transition
is of first order, then a geometrically structured
extended region will form 
in the superdense interior of the neutron star, where the
two phases are in 
equilibrium \cite{nkg91}. 
The reason for this richness in structure is that a
neutron star has 
two {\it globally} conserved charges, baryon number and
electric charge, and
two chemical potentials associated with these charges.  
Previous studies using the Maxwell construction could only
ensure one 
chemical potential was common in the two phases, whereas
the general phase 
equilibrium criteria by Gibbs \cite{gibbs} ensure
thermodynamical equilibrium
for a system with any number of chemical potentials.
The consequences are that the system is not locally charge
neutral and a 
competition between Coulomb and surface energies are
responsible for the 
geometrical structures. Moreover, the common pressure will
vary with the
proportion of the phases, and thus create an extended
mixed phase region 
with structure in the neutron star.

For a first order deconfinement transition, studies of the
detailed 
crystalline structure of the mixed phase region have
always been hindered 
by the lack of a single good model describing both phases.
This is in contrast to the first order transition to a
kaon condensed phase
described in \cite{glsb98,glsb99}, where both the normal
nuclear matter phase 
and the 
kaon condensed phase are described by the same
relativistic mean-field model,
which allow us to calculate the profiles of all important
quantities across
the interface. From these profiles the Coulomb energy and
the surface tension
can be found, where, especially for the latter, only
educated
guesses were previously possible.  
A condensate consisting of negatively charged kaons is
favored in neutron 
stars because they, contrary to other kaon types, can
replace electrons as 
neutralizing agents
\cite{gl85,kaplan,brown,thor,fujii,li}.

Detailed knowledge of the structure of a possible mixed
phase region at 
densities above saturation is important, irrespective of
which first order 
phase transition is responsible for it, as it may have
important consequences 
for transport and superfluid properties and rotation in
the form of r-mode 
instabilities, non-canonical values of the braking index,
and glitch 
phenomena in pulsars \cite{glbook,weber,reddy,madsen}. 

In the present paper we calculate the surface properties,
i.e., the surface 
tension and the curvature coefficient, for a semi-infinite
slab of normal 
nuclear matter in phase equilibrium with a semi-infinite
slab of kaon 
condensed matter and show the resulting crystalline
structure in the central 
part of the neutron star. 
Since we assure compliance with Gibbs phase equilibrium
criteria, the two
phases cannot be separately charge neutral, though overall
the net charge 
vanishes. Thus for a infinite system
we cannot explicitly take Coulomb interactions into
account in the Lagrangian.
However, it turns out that the typical radii of the
geometrical structures
are smaller than the Debye screening lengths of about 10
fm 
\cite{pethick93}, and therefore it is a reasonable first
approximation to 
ignore this effect in the calculation of the surface
tension. 

Section II contains a description of the non-uniform
relativistic mean-field 
model used to describe both phases. In Sec.\ III the
surface properties are 
described, whereas the consequences for the crystalline
structure in a 
neutron star are illustrated in Sec.\ IV. Finally our
results are summarized 
in Sec.\ V.

\section{NON-UNIFORM RELATIVISTIC MEAN-FIELD MODEL}
The relativistic mean-field model \cite{waletal,boguta}
used here is 
described in detail in
\cite{glsb99}. In this model the nucleon and kaon
interactions are treated 
on an equal footing which means both couple to the scalar
meson $\sigma$, the 
vector meson {\boldmath $\omega$}, and the isovector meson
{\boldmath $\rho$}, denoted $\sigma$, $V_\mu$, and
$\vec{R}_\mu$, respectively.
The Lagrangian for the nucleons is given by 
\begin{eqnarray}
{\cal L}_N \!=&& \overline{\Psi}_N \! \left( i\gamma^\mu
\partial_\mu-m_N^\ast
-g_{\omega N}\gamma^\mu V_\mu -g_{\rho N}\gamma^\mu
\vec{\tau}_N\cdot \vec{R}_\mu \!\right)\! \Psi_N \nonumber
\\
&&{} +\frac{1}{2}\partial_\mu \sigma
\partial^\mu\sigma-\frac{1}{2}m_\sigma^2\sigma^2-U(\sigma)-\frac{1}{4}
V_{\mu\nu}V^{\mu\nu} \nonumber \\ 
&&{} +\frac{1}{2}m_\omega^2V_\mu
V^\mu-\frac{1}{4}\vec{R}_{\mu\nu}
\cdot\vec{R}^{\mu\nu}+\frac{1}{2}m_\rho^2\vec{R}_\mu \cdot
\vec{R}^\mu,
\end{eqnarray}
where $ m_N^\ast\equiv m_N-g_{\sigma N}\sigma $, $
V_{\mu\nu}\equiv 
\partial_\mu V_\nu - \partial_\nu V_\mu, $  
$ \vec{R}_{\mu\nu}\equiv \partial_\mu \vec{R}_\nu
-\partial_\nu \vec{R}_\mu $,
and the scalar self-interactions are on the form
$U(\sigma)=
(1/3)bm_N(g_{\sigma N}\sigma)^3 + (1/4)c(g_{\sigma
N}\sigma)^4$, where $b$ and $c$ are constants.
$\Psi_N$ is the nucleon field operator and $\vec{\tau}_N$
is the isospin 
operator.

The kaon Lagrangian is given by
\begin{equation}
{\cal L}_K ={\cal D}_\mu^\ast K^\ast {\cal D}^\mu K
-m_K^{\ast 2} K^\ast K, 
\end{equation}
where $ {\cal D}_\mu\equiv \partial_\mu+ig_{\omega
N}V_\mu+ig_{\rho N}
\vec{\tau}_K
\cdot \vec{R}_\mu $ and $ m_K^\ast \equiv m_K-g_{\sigma
K}\sigma$. 
$K$ denotes the $K^-$ field and $\vec{\tau}_K$ the kaon
isospin operator.

In the mean-field approximation only the 0'th component of
the vector fields 
and the isospin 3-component of the isovector field have
finite mean values. 
The equation of motion for the kaon is
\begin{equation}
\left( {\cal D}_\mu{\cal D}^\mu+m_K^{\ast 2} \right)K=0.
\end{equation}
In a semi-infinite system the kaon field is only
translational
invariant in the transverse direction ${\bf z}_\perp$ and
can then be written 
as
\begin{equation}
 K=\phi(z) e^{i\left(\omega_K+{\bf k}_\perp\cdot {\bf
z}_\perp\right)},
\end{equation}
where ${\bf k}_\perp$ is the transverse momentum,
$\omega_K$ is the in-medium
kaon energy, and $\phi(z)$ is the kaon amplitude.

For s-wave condensation, i.e. ${\bf k}=0$, the kaon field
equation is given by 
\begin{equation}
\label{eqmkaon}
  \nabla^2 \phi +\left( \left(\omega_K+g_{\omega
K}V_0+g_{\rho K}
  R_{0,3} \right)^2-m^{\ast~ 2}_K \right)\phi=0.
\end{equation}
In the semi-infinite case $\nabla^2 =d^2/dz^2$, where $z$
is the direction 
perpendicular to the surface.

The kaon density is
\begin{equation}
\label{rhokaon}
 \rho_K=2\left( \omega_K+g_{\omega K}V_0+g_{\rho
K}R_{0,3}\right)
 K^\ast K,
\end{equation} 
and only for an infinite system $\rho_K=2m_K^\ast K^\ast
K$ as seen from Eq.\
(\ref{eqmkaon}). 

The equations of motion for the position dependent meson
fields become
\begin{eqnarray}
\label{eqmsig}
 \nabla^2 \sigma - m_\sigma^2 \sigma = && g_{\sigma
N}\left( -\rho_s
 +bm_N(g_{\sigma N}\sigma)^2+c(g_{\sigma N}
\sigma)^3\right) \nonumber \\
 &&{} -2g_{\sigma K}m^\ast_K K^\ast K,
\end{eqnarray}
\begin{equation}
\label{eqmome}
 \nabla^2 V_0 -m_\omega^2 V_0 = -g_{\omega
N}(\rho_n+\rho_p)+
 g_{\omega K}\rho_K,
\end{equation}
and
\begin{equation}
\label{eqmrho}
 \nabla^2 R_{0,3} - m_\rho^2 R_{0,3} = -g_{\rho
N}(\rho_p-\rho_n)
 +g_{\rho K}\rho_K,
\end{equation}
where the neutron and proton densities, $\rho_n$ and
$\rho_p$, and the 
scalar density $\rho_s$ are calculated in the Thomas-Fermi
(local density) 
approximation. 
\begin{equation}
\label{density}
\rho_i=\frac{1}{3\pi^2} k_i^3,   \hspace{2cm} i=n,p
\end{equation}
and 
\begin{eqnarray}
\label{scalardens}
\rho_s= \frac{m_N^{\ast}}{2\pi^2} && 
 \sum_{i=n,p} \left( k_i\sqrt{k_i^2+m_N^{\ast2}} \right.
\nonumber \\
&&{} - m_N^{\ast2} 
\left.
\ln\left(\frac{k_i+\sqrt{k_i^2+m_N^{\ast2}}}{m_N^{\ast}}\right)\right).
\end{eqnarray}
Expressions for the local Fermi momenta $k_i$ are obtained
from the Dirac 
equation for the nucleons
\begin{equation}
\label{dirac}
\mu_i =g_{\omega N}V_0\mp g_{\rho N} R_{0,3}
+\sqrt{k_i^2+m_N^{\ast2}},
\end{equation}
where the upper sign is used for neutrons. 
The electron chemical potential $\mu_e=\mu_n-\mu_p$.
The kaon amplitude is zero unless the condition
$\omega_K=\mu_{K^-}=\mu_e$
is fulfilled.

The model parameters $g_{iN}/m_i$
($i=\sigma,\omega,\rho$), $b$, and $c$ can be 
algebraically 
determined from the five bulk properties of nuclear
matter, which we take as:
$E/A=-16.3$ MeV, $\rho_0=0.153$ fm$^{-3}$, $a_{sym}=32.5$
MeV, $K= 240$ MeV,
and $m^{\ast}/m_N=0.78$. Because of the Laplacian terms in
the meson field 
equations explicit values for the masses are required. We
use for the 
{\boldmath $\omega$} and {\boldmath $\rho$} their rest
masses 
$m_{\omega}=782$ MeV and $m_{\rho}=768$ MeV, while the
$\sigma$ mass is 
determined from the surface properties of symmetric
nuclear matter as described
in \cite{boguta,vinas93}, $m_{\sigma}=390$ MeV.
The kaon coupling constants for the vector mesons are
determined from the 
quark and isospin counting rule,
\begin{equation}
 g_{\omega K}=g_{\omega N}/3 \hspace{0.7cm} \hbox{and}
\hspace{0.7cm} 
g_{\rho K}=g_{\rho N}, 
\end{equation}
and the scalar coupling constant is fixed to the optical
potential of the 
$K^-$ at $\rho_0$ from
\begin{equation}
 U_K(\rho_0)=-g_{\sigma K}\sigma(\rho_0)-g_{\omega
K}V_0(\rho_0),
\end{equation}
where we for the optical potential use $U_K(\rho_0)=-120$
MeV.

The Gibbs conditions for an infinite system of a kaon
condensed phase and 
a normal nuclear matter phase to be in thermodynamical
equilibrium at zero 
temperature are
\begin{eqnarray}
\mu_{i,N}&=&\mu_{i,K} \hspace{2cm} i=n,e \nonumber \\
P_N(\mu_n,\mu_e)&=&P_K(\mu_n,\mu_e),
\end{eqnarray}
where $P_N$ and $P_K$ are the pressures of the normal
nuclear matter phase 
and the kaon condensed phase, respectively.
These conditions combined with the global condition of
electric charge 
conservation,
\begin{equation}
q_{total}=(1-\chi)q_N(\mu_n,\mu_e)+\chi q_K(\mu_n,\mu_e),
\end{equation}
where $q$ denotes the charge of the corresponding phase,
and the field equations (without the Laplacian terms)
allow us to solve for 
the bulk values of the fields, 
densities, and chemical potentials in each phase for any
volume fraction of 
kaon phase $\chi$ between zero and one.

\begin{figure}[ht]
\vspace{-.2in}
\begin{center}
\leavevmode
\psfig{figure=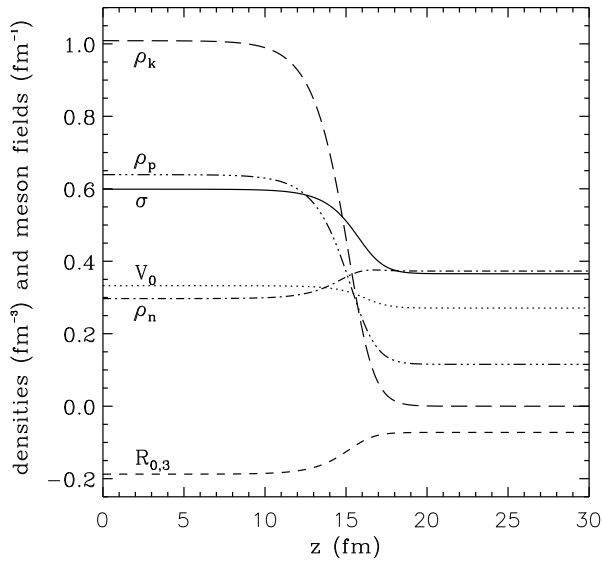,width=3.4in,height=3.19in}
\parbox[t]{5.5in} { \caption { \label{fig1} Profiles of
the meson fields 
$\sigma$, $V_0$, and $R_{0,3}$ and the neutron, proton,
and kaon densities, 
$\rho_n$, $\rho_p$, and $\rho_k$, respectively, across the
interface.
The boundary conditions at $z=0$ and 30 fm correspond to
$\chi=0.107$.}}
\end{center}
\end{figure}

In order to calculate the surface tension between the kaon
condensed phase and
the normal nuclear matter phase, the profiles of the
fields and densities
have to be determined across the interface between
semi-infinite slabs of
each phase. This is done by simultaneously solving the
four coupled 
differential equations for the $K^-$ and the meson fields
Eqs.\ 
(\ref{eqmkaon},\ref{eqmsig}-\ref{eqmrho}) 
through a relaxation procedure, where initial guesses for
the 
different profiles are relaxed to their equilibrium
values. The boundary 
conditions at $\pm \infty$ are provided by the bulk values
of the kaon 
amplitude and meson fields for each phase at a fixed
$\chi$.
In practice a 30 fm region with the interface placed
approximately in the center is sufficient to fulfill the
boundary conditions,
see Fig.\ \ref{fig1}.
%\noindent

It is a surprisingly good approximation and a significant
simplification to 
neglect the 
Laplacian terms in the meson field equations, and only
keeping the term in the
kaon field equation. Differentiating twice brings the kaon
equation to 
the form $d^2\phi/dz^2 =X/$
$\left(\left(\omega_K+g_{\omega K}V_0
+g_{\rho K} R_{0,3} \right)^2-m^{\ast~ 2}_K \right)$. 
The denominator vanishes in the bulk kaon condensed phase
in contrast to 
the meson field equations, which take the form 
$d^2\sigma /dz^2 \simeq Y/m^2_{\sigma}$ ($X$ and $Y$
denote two complicated
functions). Thus the Laplacian term is
generally most important for the kaon field equation.
In Ref.\ \cite{reddy} the approximation was used to
calculate the weak 
charge of droplets of kaon condensed phase to study
effects on the neutrino 
opacity. 
We do both calculations and use the results from the
approximative procedure
as initial guess for the solution to the full set of four
differential 
equations. This procedure ensures rapid convergence.

\section{Surface Properties}
In the transition zone between two phases the pressure
tensor is no longer 
isotropic as it is for the homogeneous bulk phases. For a
plane interface 
with the normal in the $z-$direction, the pressure tensor
can be split into 
a normal and a tangential component. The normal component
of the pressure $P$
stays constant across the interface, cf.\ to Gibbs phase
equilibrium 
conditions, whereas the tangential pressure component
$P_T(z)$ changes 
as function of $z$ across the interface. 
The mechanical definition of the surface tension
$\sigma_G$ of a plane 
interface is \cite{hirsch-ono}
\begin{equation}
 \sigma_G=\int^{\infty}_{-\infty} (P-P_T(z))dz,
\end{equation}
where $P_T(-\infty)=P_T(\infty)=P$.
In nuclear physics this definition of the surface tension
is denoted Gibbs 
definition \cite{myers85,vinas98}.
 
Because we have assumed the leptonic species are
homogeneously distributed 
throughout the system, they do not contribute to the
surface 
tension. The pressure of the hadronic species $P_H$ is
given by
\begin{equation}
 P_H=\rho \frac{\partial \epsilon_H}{\partial \rho}
-\epsilon_H 
    =\mu_n\rho_n+\mu_p\rho_p+\mu_e\rho_k-\epsilon_H.
\end{equation}
Therefore the surface tension of a plane interface between
a kaon condensed 
phase and a normal nuclear matter phase can be written as 
\begin{eqnarray}
\label{sigma}
 \sigma_G=\int^{\infty}_{-\infty}&dz&\left(
\epsilon_H(z)-\epsilon_{H,N} 
 -\mu_n\left( \rho_n(z)-\rho_{n,N}\right)  \right.
\nonumber \\
 &&{} -\left. \mu_p\left(\rho_p(z)-\rho_{p,N}\right)
 -\mu_e\rho_k(z)\right),
\end{eqnarray}
where the $z$ dependent quantities at minus infinity take
the bulk values of 
the kaon condensed phase and at plus infinity take the
bulk values of the 
normal nuclear matter phase. The energy density of the
hadrons across the 
interface is given by
\begin{eqnarray}
 &&\epsilon_H(z)=\frac{1}{2}\left( (\nabla
\sigma)^2+m_\sigma^2 \sigma^2 \right)
 +\frac{1}{2}\left( (\nabla V_0)^2+m_\omega^2 V_0^2
\right) 
 \nonumber \\
 &&+\frac{1}{2}\left( (\nabla R_{0,3})^2\!+m_\rho^2
R_{0,3}^2 \right)
 \!+\!2\left(\omega_K\!+g_{\omega K}V_0+g_{\rho K} R_{0,3}
\right)^2
 \!\phi^2       \nonumber \\
 &&{}+U(\sigma) +\sum_{i=n,p} \frac{1}{\pi^2} 
 \int_0^{k_i}\sqrt{k^2+m_N^{\ast2}}\,k^2\;dk.
\end{eqnarray}
Notice, if the Laplacian terms are neglected except for
the kaon field, the 
only consequence is that the squared gradient terms of
$V_0$ and $R_{0,3}$ 
change sign.

The curvature coefficient is in Gibbs definition given by 
\cite{myers69,vinas98}
\begin{eqnarray}
\label{gamma}
 \gamma_G &=&\int_{-\infty}^{\infty} dz(z-z_0)\left(
\epsilon_H(z)-
 \epsilon_{H,N} -\mu_n\left( \rho_n(z)-\rho_{n,N}\right)
\right.  \nonumber \\
 &&{} -\left. \mu_p\left(\rho_p(z)-\rho_{p,N}\right)
 -\mu_e\rho_k(z)\right),
\end{eqnarray}
where the surface location, i.e., the position of the
equivalent sharp 
surface, 
\begin{equation}
 z_0=\frac{\int_{-\infty}^\infty z \rho^\prime(z)\; dz} 
     {\int_{-\infty}^\infty \rho^\prime(z)\; dz},
\end{equation}
and the prime denotes the derivative with respect to $z$. 
\begin{figure}[ht]
\vspace{-.2in}
\begin{center}
\leavevmode
\psfig{figure=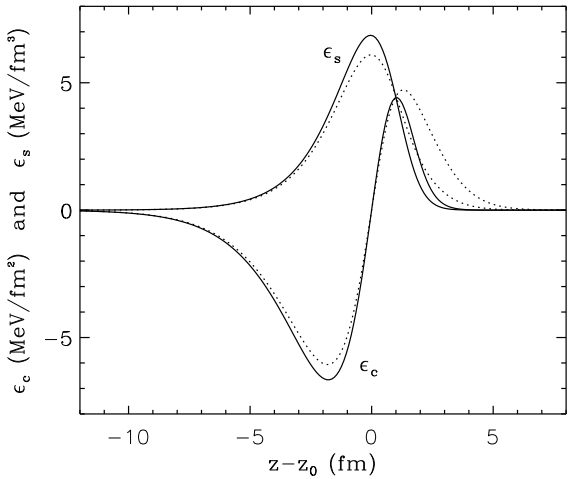,width=3.4in,height=2.76in}
\parbox[t]{5.5in} { \caption { \label{fig2} The surface
tension and curvature 
coefficient densities, $\epsilon_s$ and $\epsilon_c$,
respectively, across the
interface. The solid curves are for the approximative
calculation and the 
dotted curves for the exact calculation.}}
\end{center}
\end{figure}
%\noindent
In Fig.\ \ref{fig2} the surface tension and curvature
coefficient densities 
(i.e., the 
integrands of Eqs.\ (\ref{sigma}) and (\ref{gamma})),
$\epsilon_s$
and $\epsilon_c$, respectively, are
plotted both for the full set of differential equations
and for the 
approximation with only a differential equation for the
kaon field.
The differences between the two surface tension density
profiles are small. 
The exact calculation result in a slightly broader and 
more symmetric profile with a smaller peak value than the
approximation. 
The curvature coefficient density
$\epsilon_c=(z-z_0)\epsilon_s$, and is 
therefore naturally more sensitive to the exact form,
especially the width,
of $\epsilon_s$. This is also seen from the plot. 
The densities shown in Fig.\ \ref{fig2} are for bulk
parameters corresponding 
to $\chi=0.107$. For $\chi$ increasing from 0 to 1 the
width of $\epsilon_s$
increases from about 4 fm to 7 fm, while the peak value
decreases. 

\begin{figure}[ht]
\vspace{-.2in}
\begin{center}
\leavevmode
\psfig{figure=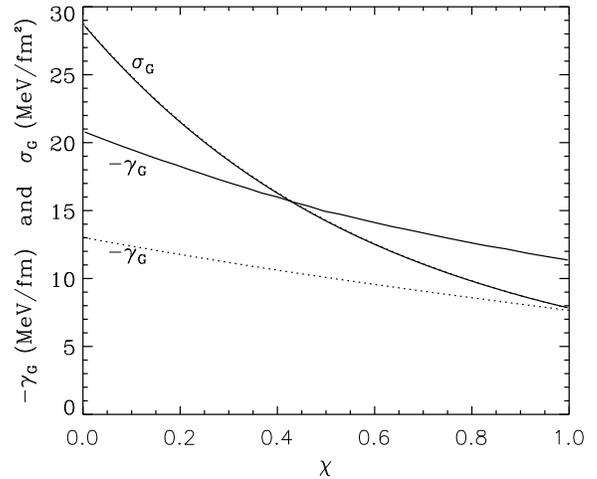,width=3.4in,height=2.76in}
\parbox[t]{5.5in} { \caption { \label{fig3} The surface
tension $\sigma_G$ 
and curvature coefficient $\gamma_G$ as a function of the
volume fraction 
of kaon condensed phase $\chi$.
The solid curves are for the approximative calculation and
the
dotted curves for the exact calculation.}}
\end{center}
\end{figure}
%\noindent
The overall consequences for the surface tension and
curvature
coefficient are shown in Fig.\ \ref{fig3}, where
$\sigma_G$ and $-\gamma_G$ 
have been plotted as a function of $\chi$. Notice, the
curvature coefficient 
is negative. 
The surface tension decreases monotonically with
increasing volume fraction 
of the kaon phase, whereas the curvature coefficient
increases. 
The values of $\sigma_G$ from the exact and approximative
procedure deviate 
less than 0.4\%. Thus the two curves lie on top of each
other in this plot. 
This is in contrast to the curvature, where the difference
is about 33\%  
between the exact and approximative calculation. 
Furthermore, the absolute value of the curvature
coefficient is comparable to 
the surface tension. This means the curvature energy is
not negligible 
compared to the surface energy. But of course even higher
order effects may 
mask the effect of the curvature energy as in the nuclear
mass formula, where
the apparent absence of a curvature term seem to be caused
by higher order
effects \cite{myers96}.
For a spherical droplet of kaon phase with radius $R=5$ fm
in a system where 
$\chi \rightarrow 0$, the surface energy is reduced about 
$2\gamma_G/(\sigma_GR)=18$ \%,
whereas the surface energy {\it increases}, since the
curvature is a signed 
quantity, about 40 \% for a spherical bubble of normal
nuclear matter phase in 
the $\chi \rightarrow 1$ limit. We have here assumed, that
both $\sigma_G$ 
and $\gamma_G$ calculated from a plane interface remain
unaltered. 

\begin{figure}[ht]
\vspace{-.2in}
\begin{center}
\leavevmode
\psfig{figure=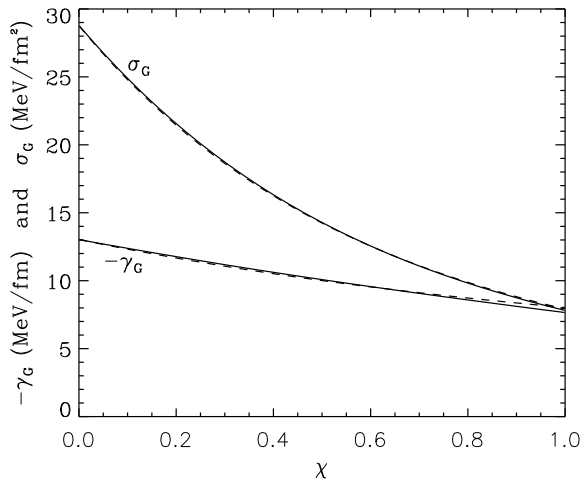,width=3.4in,height=2.76in}
\parbox[t]{5.5in} { \caption { \label{fig4} The surface
tension $\sigma_G$
and curvature coefficient $\gamma_G$, solid curves,
compared with the fits
given in Eq.\ (\ref{fit}), dashed curves, as a function of
the volume fraction
of kaon condensed phase $\chi$.}}
\end{center}
\end{figure}

In Fig.\ \ref{fig4} the exact calculation of $\sigma_G$
and $-\gamma_G$ as a 
function of $\chi$ have been compared with the fits 
\begin{eqnarray}
\label{fit}
\sigma_{fit}&=&\left(0.00786\,\frac{\epsilon_K-\epsilon_N}{\mbox{MeV/fm$^3$}}
 \right)^2\:\mbox{MeV/fm}^2 \nonumber \\
\gamma_{fit}&=&-0.0976\left(\frac{\epsilon_K-\epsilon_N}{\mbox{MeV/fm$^3$}}
 \right)^{3/4}\:\mbox{MeV/fm},
\end{eqnarray}
where the energy densities are in units of MeV/fm$^3$.  
The fit deviate less than 5\% for the curvature
coefficient and less than 0.5\%
for the surface tension over the region where
$\epsilon_K-\epsilon_N$ decreases
from 682 MeV to 358 MeV, i.e., about 50\%. 
It is not exactly trivial to understand why the fits,
especially for the 
surface tension, are so good. For the surface tension,
Eq.\ (\ref{fit}) is 
equivalent to saying
\begin{equation}
 (\epsilon(z)-\epsilon_N)\frac{d \epsilon}{d z} \propto 
 P_N-P(z),
\end{equation}
whenever $\epsilon_s=P_N-P(z)$ is not negligible. This is
indeed the case as 
long as $d^2\epsilon/dz^2$ is negative but certainly not
otherwise. 
We have not pursued the issue any further, but 
suspect this may be a generic property of phase equilibria
described in  
mean-field theory in the Thomas-Fermi approximation. 
%\noindent

\section{CRYSTALLINE STRUCTURE IN NEUTRON STARS}

\begin{figure}[ht]
\vspace{-.2in}
\begin{center}
\leavevmode
\psfig{figure=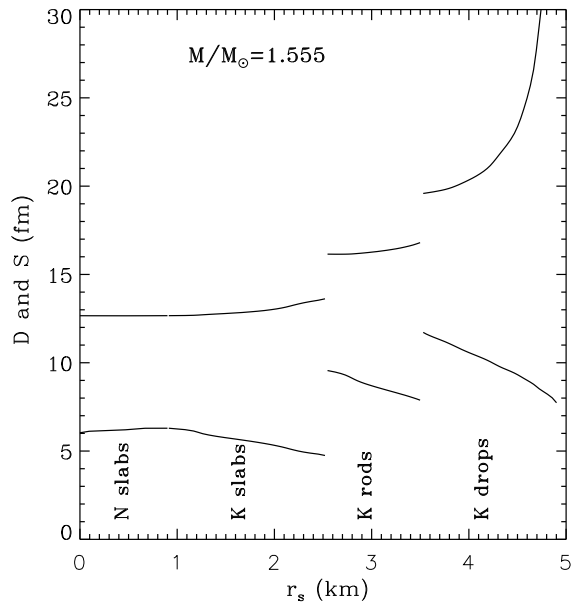,width=3.4in,height=3.4in}
\parbox[t]{5.5in} { \caption { \label{fig5} The diameter
$D$ (lower curve)
of the geometries of the rarer phase immersed in the
dominant phase and the 
spacing $S$ between the geometries as a function of the
radial coordinate 
$r_s$.}}
\end{center}
\end{figure}

In Ref.\ \cite{glsb99} the surface tension was assumed to
be proportional 
to the difference in energy density between the two phases
and the constant 
chosen rather arbitrarily, but in a way that ensured the
sum of surface and
Coulomb energies was always much smaller than the bulk
energy. 
Furthermore, the Coulomb energy was, as usual, calculated
under the assumption
that the phase boundary is sharp. Figure 18 of that paper
show the diameter
$D$ and the spacing $S$ of the geometrical structures as a
function of the 
radial coordinate for a star at the mass limit. 
As we have now shown, the surface tension is proportional
to the difference 
in energy density squared, and it is between a factor of
2-5 smaller 
than what was guessed at in \cite{glsb99}. 
Knowledge of the density profiles of the charged particles
across the plane
interface allow us to calculate the Coulomb energy for a
soft boundary. 
It turns out that the Coulomb energy is reduced less than
15\% for 
geometries with the typical small sizes encountered here.
This translate into 
about a 5\% reduction of the sum of surface and Coulomb
energies. Therefore 
it is a good approximation to treat the phase boundary as
being sharp with 
respect to calculating the Coulomb energy. 
Figure \ref{fig5} is an updated version of Fig.\ 18 of
Ref.\ \cite{glsb99}. 
(Details of this type of calculation can be found in
\cite{pei}.) 
The differences between the two figures are minor. 
Overall the sizes of the geometrical structures just
decrease about 30\%.
This is due to the fact, that to first 
approximation the location of the transition from one
geometry to another is 
independent of $\sigma_G$. Furthermore, the sum of surface
and Coulomb 
energies scale only as $\sigma_G^{1/3}$. 
%\noindent

We have ignored the curvature term completely in these
calculations, even 
though it is straight forward to include the curvature
energy, see 
\cite{glmbc}. However, if the curvature energy is
negative, as it is for 
kaon droplets, the sum of finite size and Coulomb energies
will not have a
local minimum but will decrease to minus infinity with
decreasing size of the 
structure. This is of course unphysical and higher order
effects will prevent 
it from happening. These complicated higher order effects
may, as previously 
mentioned, even to some extent mask the presence of a
curvature term. 
Moreover, the curvature coefficient is very sensitive to
the exact profile 
of the phase boundary as we have seen, so its real value
is uncertain for 
the rather small structures encountered here. 
Therefore we choose to consider the value of the curvature
coefficient as 
some measure for the amount of uncertainty in the surface
tension calculation,
but not more than that.

\section{SUMMARY}

Little is known about the equation of state for nuclear
matter above
saturation density, but it is expected that at least a
phase transition to 
deconfined quark matter happens at a few times the
saturation density.
However, this phase transition may not be the only one
encountered in 
neutron stars. A transition to a phase with a $K^-$
condensate is also a 
possibility.
If any of these phase transitions is of first order, a
neutron star will have 
a mixed phase region in its dense interior, which is very
likely to have some 
observable consequences. 

For the first time the surface properties in the interface
between normal 
nuclear matter and kaon condensed matter have been
calculated, which makes it 
possible to study this mixed phase region in greater
detail. 
Our calculations are only a first approximation, there are
a number of
complicating aspects, which we have ignored or only
treated approximately -
e.g., explicit consideration of the Coulomb field which
results in screening 
effects;
the validity of the assumption that the surface tension is
the same for 
semi-infinite slabs as it is for small slabs, rods, and
drops; and the 
importance of the curvature and even higher order terms
compared to the 
surface term. 
Concerning the first two points, we can generally say that
screening effects
will reduce the Coulomb energy, likewise the surface
energy will decrease 
due to the decrease in the surface tension when the system
is squeezed. The 
reason for the latter is that the surface region will
dominate a small system,
so that only $\epsilon_s$ and not $d\epsilon_s/dz$ is zero
at the boundaries. 
For example a slab which is 10 fm thick and with a volume
fraction of kaon 
condensed phase of about 0.4, the surface tension is
reduced about 12\% 
compared to the surface tension of the infinite system (30
fm thick), while 
the absolute value of the curvature coefficient drops
about a factor of three. 
These two effects pull in opposite directions with regard
to the size of the 
geometry which minimize the sum of Coulomb and surface
energies, thus the 
overall effect on the size is expected to be only minor.

We have taken another step towards a better understanding
of the mixed phase 
region for a first order transition involving a kaon
condensate. 
We do, however, realize there is still much room for
improvement. 

\section*{ACKNOWLEDGMENTS}
We wish to thank M.\ Centelles and X.\ Vi\~{n}as for
providing their 
relaxation codes to us. M.B.C.\ acknowledges financial
support by the 
Carlsberg Foundation. \doe

\end{document}